\documentclass{article}%
\usepackage{amsmath}
\usepackage{amsfonts}
\usepackage{amssymb}
\usepackage{graphicx}%
\providecommand{\U}[1]{\protect\rule{.1in}{.1in}}

\begin{document}

\begin{center}
\textbf{Remarks On The Spherical Scalar Field Halo In Galaxies}

\bigskip

Kamal K. Nandi$^{1,2,3,a}$, Ildar Valitov$^{2,b}$ and Nail G. Migranov$^{3,c}$

$^{1}$Department of Mathematics, University of North Bengal, Siliguri 734 013, India

$^{2}$Department of Theoretical Physics, Sterlitamak State Pedagogical
Academy, Sterlitamak 453103, Russia

$^{3}$Joint Research Laboratory, Bashkir State Pedagogical University, Ufa
450000, Russia

\bigskip

$^{a}$E-mail: kamalnandi1952@yahoo.co.in

$^{b}$E-mail: diamond921@yandex.ru

$^{c}$E-mail: ufangm@yahoo.co.uk

\bigskip

\textbf{Abstract}
\end{center}

Matos, Guzm\'{a}n and Nu\~{n}ez proposed a model for the galactic halo within
the framework of scalar field theory. We argue that an analysis involving the
full metric can reveal the true physical nature of the halo only when a
certain condition is maintained. We fix that condition and also calculate its
impact on observable parameters of the model.

\begin{center}
------------------------------------------------------------
\end{center}

\bigskip

One of the outstanding problems in modern astrophysics is the problem of dark
matter which is invoked as an explanation for the observed flat rotation
curves in the galactic halo. Doppler emissions from stable circular orbits of
neutral hydrogen clouds in the halo allow the measurement of tangential
velocity $v_{tg}(r)$ of the clouds treated as probe particles. According to
Newton's laws, centrifugal acceleration $v_{tg}^{2}/r$ should balance the
gravitational attraction $GM(r)/r^{2}$, which immediately gives $v_{tg}%
^{2}=GM(r)/r$. That is, one would expect a fall-off of $v_{tg}^{2}(r)$ with
$r$. However, observations indicate that this is not the case:$\ v_{tg}$
approximately levels off with $r$ in the halo region. The only way to
interpret this result of observation is to accept that the mass $M(r)$
increases linearly with distance $r$. Luminous mass distribution in the galaxy
does not follow this behavior. Hence the hypothesis that there must be huge
amounts of nonluminous matter hidden in the halo. This unseen matter is given
a technical name dark matter.

Despite the fact that the exact nature of dark matter is as yet unknown,
several analytic halo models exist in the literature including those provided
by scalar-tensor theories (see for instance [1]). In particular, the scalar
field model first proposed by Matos, Guzm\'{a}n and Nu\~{n}ez [2] has received
considerable attention. It is important to note that the authors primarily
constructed an exact solution of Einstein's field equations sourced by a
scalar field that provides a density profile of $1/r^{2}$ together with other
appealing features of the metric functions. As a particular application, they
sketched a plausible interpretation of the halo dark matter problem. The
problem being important in itself, we think that the interesting relativistic
central feature of the solution, namely, a non-Newtonian halo, must be well
grounded. The purpose of the present Brief Report is to fix the condition
under which it is possible. In addition, we work out its impact on observable parameters.

It is to be mentioned that the solution in [2] has been criticized because of
its singular behavior at the origin [3], but this singularity is not peculiar
to that solution alone; there are other viable halo models in the literature
that also possess such singularity (see for instance [4]). While we are here
interested only in the outer reaches of the galaxy (very far from the origin),
the authors of Ref.[2] and the associated research group have addressed this
inner region singularity too. They obtained several new results under the
scalar field dark matter model in galaxies: Solution with axial symmetry
including the inner region [5], time-dependent spacetimes [6], the full
non-linear Newtonian evolution after the turn-around point [7], time evolution
of density fluctuation [8], collision properties of two structures [9] and so
on. While they obtained constraints arising out of cosmological scale
considerations, we believe that it is also useful to ascertain the constraint
arising out of the local scale, which would clarify the relativistic nature of
the spherically symmetric model of the \ halo under consideration.

Using the flat rotation curve condition [10], Matos, Guzm\'{a}n and Nu\~{n}ez
obtain the spherically symmetric static solution for the galactic halo as
follows ($G=c=1$, unless specifically restored):%
\begin{equation}
ds^{2}=-B(r)dt^{2}+A(r)dr^{2}+r^{2}(d\theta^{2}+sin^{2}\theta d\phi^{2})
\end{equation}%
\[
B(r)=B_{0}r^{l}%
\]%
\begin{equation}
A(r)=\frac{4-l^{2}}{4+D(4-l^{2})r^{-(l+2)}}%
\end{equation}%
\begin{equation}
\phi(r)=\sqrt{\frac{l}{8\pi}}\ln r+\phi_{0}%
\end{equation}%
\begin{equation}
V(r)=-\frac{1}{8\pi(2-l)r^{2}},
\end{equation}
where $D$ is an arbitrary constant of integration, $\phi$ and $V$ are the
scalar field and potential respectively. The parameter $l=2(v_{tg}/c)^{2}$,
$B_{0}>0$ is another constant. Observations of the frequency shifts in the HI
radiation show that, in the halo region, $v_{tg}/c$ is nearly constant at a
value $7\times10^{-4}$~[11]. Thus, in what follows, we take $l\sim10^{-6}$.

Note that we can rewrite $A(r)$ in the standard Schwarzschild form%
\begin{equation}
A(r)=\left[  1-\frac{2m(r)}{r}\right]  ^{-1}%
\end{equation}
which is often convenient and will be useful later while discussing the
observational parameters. Such a form has the advantage that it immediately
reveals not only the mass parameter $m(r)$ but also shows that the proper
radial length is larger than the Euclidean length because $r>2m(r)$. This
inequality, which is essential for signature protection, dictates that
$A(r)>1$. This is a crucial condition to be satisfied by any valid metric.

Now, for the sake of simplicity, Matos, Guzm\'{a}n and Nu\~{n}ez choose $D=0$,
but this is not the best choice because it makes the metric component $A<1$.
As a consequence, whatever results follow from the reduced metric should be
taken with caution. For instance, the stresses exhibit a density profile
$\rho<0$, meaning violation of Weak Energy Condition (WEC) and furthermore
lead to $\omega<-1$ (see below), meaning repulsive gravity in the halo,
contradicting observational facts. But these are actually not the true
features of their model. To see the true picture, it is necessary to calculate
the relevant quantities with $D\neq0$.

We find the density and pressure profiles in the rest frame of the fluid as%
\begin{equation}
\rho=\frac{1}{8\pi}\frac{r^{-(4+l)}[D(l^{3}+l^{2}-4l-4)+l^{2}r^{2+l}]}%
{l^{2}-4}%
\end{equation}%
\begin{equation}
\text{ }p_{r}=\frac{1}{8\pi}\frac{r^{-(4+l)}[D(l^{3}+l^{2}-4l-4)-l(4+l)r^{2+l}%
]}{l^{2}-4}%
\end{equation}%
\begin{equation}
\text{ }p_{t}=\frac{1}{8\pi}\frac{r^{-(6+l)}[D(l^{3}+l^{2}-4l-4)+l^{2}%
r^{2+l}][(r^{2}-1)l-2(r^{2}+1)]}{4(l^{2}-4)}%
\end{equation}
where $\rho$ is the energy density, $p_{r}$ is the radial pressure and $p_{t}$
are the transverse pressures.

Matos, Guzm\'{a}n and Nu\~{n}ez conclude that their model has huge pressure
over density and thus it is non-Newtonian. We wish to emphasize that the role
of non-zero value of $D$ is crucial not only for avoiding repulsive gravity
(as alluded to above) but also for arriving at a correct conclusion about the
relative strengths between pressure and density. For instance, let us take
$D=1$. In the distant halo region, we can take, typically, $r\sim100-300$
$Kpc$ and with $l\sim10^{-6}$, we find the numerical values to be $\rho
\sim10^{-9\text{ }}$ and $p_{r}\sim10^{-9}$, which means that they are of the
same order. But on the other hand, $p_{r}+2p_{t}\sim10^{-11}\Rightarrow
p_{r}+2p_{t}\sim10^{-2}\rho$, which indicates that total pressure is roughly
one hundred times \textit{less} than the density. However, if we take
$D=10^{-5}$, we find that $p_{r}+2p_{t}\sim10^{3}\rho$. If we keep on
decreasing the value of $D$ further (but never exactly to zero for reasons
stated above), we see that the total pressure dominates more and more over
density reinforcing the non-Newtonian nature.

The next question is: How far can we go on decreasing $D$? We notice the
following interesting scenario. When $D=10^{-7}$, we find $p_{r}%
+2p_{t}=9\times10^{5}\rho$, which leads to $\omega=\frac{p_{r}+2p_{t}}{3\rho
}=3\times10^{5}$ (attractive gravity) as shown in Fig.1. This is the extreme
possible non-Newtonian halo in the scalar field model under consideration. The
reason is this. If $D=10^{-8}$, we find that $\omega>0$ up to $r=r_{0}=200$
$Kpc$ (attractive gravity) and becomes $\omega<-1$ after $r=r_{0}$ (repulsive
gravity). At $r=r_{0}$, there is a singularity in $\omega$. This value of $D$
represents a transition from attraction to repulsion as shown in Fig.2. When
$D\leq10^{-9}$, we find that $\rho<0$, $\omega<-1$ (repulsive gravity), which
share the woes that follow also from the choice $D=0$ (Figs.3 and 4). These
show that we can not decrease $D$ below $10^{-7}$, that is, we must have
$D\geq10^{-7}$. This is the condition that must be maintained in order to have
a non-Newtonian halo.

The pressures are anisotropic, as is evident from Eqs.(7,8), which is a good
feature of the solution from the point of view of exterior matching. Note that
the solution can not be matched to the Schwarzschild exterior metric at the
boundary of the halo if the pressures were isotropic [12]. It can be further
verified that $\rho>0$, $\rho+p_{r}>0$, $\rho+p_{r}+2p_{t}>0$ for
$D\geq10^{-7}$; so we can say that the halo matter is not exotic because the
standard energy conditions are satisfied everywhere. Therefore, we expect
attractive gravity in the halo. To confirm it, we follow the prescription by
Lynden-Bell, Katz and Bi\v{c}\'{a}k [13], and find that the total
gravitational energy is indeed negative:
\begin{equation}
E_{G}=4\pi\int_{r_{1}}^{r_{2}}[1-A^{\frac{1}{2}}]\rho r^{2}dr<0,
\end{equation}
due to the fact that $\rho>0$, $1-A^{\frac{1}{2}}<0$ and $r_{2}>r_{1}$. This
prescription has been very useful in the case of scalar field wormholes too [14-16].

Certainly, the scalar field model corresponding to $D=10^{-7}$ is highly
non-Newtonian because $p_{r}+2p_{t}\sim10^{6}\rho$. As a result, a purely
Newtonian definition of mass, viz., $M(r)=4\pi\int\rho r^{2}dr$ does not
apply. However, incorporating the pressure contribution, the dynamical mass in
the first post Newtonian order becomes
\begin{equation}
M_{pN}(r)=4\pi\int(\rho+p_{r}+2p_{t})r^{2}dr=10^{6}M(r),
\end{equation}
which clearly reflects the non-Newtonian nature of the model in terms of masses.

We next focus on the observable parameters expected in this non-Newtonian
halo. Whatever be the analytic model for it, there must be a way to contrast
its predictions with actual measurements. The key point is that one does not
directly measure the metric functions but indirectly measures gravitational
potentials and masses from rotation curve and lensing observations. Faber and
Visser [17] have shown how, in the first post Newtonian approximation, the
combined measurements of rotation curves and gravitational lensing allow
inferences about the mass and pressure profile of the galactic halo as well as
its equation of state.

The usual techniques for obtaining the potential for rotation curve (RC)
measurements yield a pseudo-potential (See Ref.[17] for details):%
\begin{equation}
\Phi_{\text{RC}}=\Phi\neq\Phi_{N},
\end{equation}
where $\Phi_{N}$ is the Newtonian potential, $\Phi=\frac{1}{2}\ln B$ and a
pseudo-mass
\begin{equation}
m_{\text{RC}}=r^{2}\Phi^{\prime}(r)\approx4\pi\int(\rho+p_{r}+2p_{t})r^{2}dr.
\end{equation}
Faber and Visser also define the lensing pseudo-potential as
\begin{equation}
\Phi_{\text{lens}}=\frac{\Phi(r)}{2}+\frac{1}{2}\int\frac{m(r)}{r^{2}}dr.
\end{equation}
and a pseudo-mass $m_{\text{lens}}$ obtained from lensing measurements as%

\begin{equation}
m_{\text{lens}}=\frac{1}{2}r^{2}\Phi^{\prime}(r)+\frac{1}{2}m(r).
\end{equation}
The first order approximations of Einstein's equations yield%
\begin{equation}
\rho(r)\approx\frac{1}{4\pi r^{2}}[2m_{\text{lens}}^{\prime}(r)-m_{\text{RC}%
}^{\prime}(r)]
\end{equation}%
\begin{equation}
4\pi r^{2}(p_{r}+2p_{t})\approx2[m_{\text{RC}}^{\prime}(r)-m_{\text{lens}%
}^{\prime}(r)]
\end{equation}
where the right hand sides denote pseudo-density and pseudo-pressures.
Furthermore, Faber and Visser define a dimensionless quantity%
\begin{equation}
\omega(r)=\frac{p_{r}+2p_{t}}{3\rho}\approx\frac{2}{3}\frac{m_{\text{RC}%
}^{\prime}-m_{\text{lens}}^{\prime}}{2m_{\text{lens}}^{\prime}-m_{\text{RC}%
}^{\prime}}.
\end{equation}

The pseudo quantities on the right hand side of Eqs.(11)-(17) are actual
observables from the combined measurement. If the observed pseudo profiles
reasonably match with the analytic pseudo profiles coming from \textit{a
priori} given metric functions, one can say that the solution is physically
substantiated. Otherwise, it has to be ruled out as non-viable. The impact of
a small non-zero $D$ on the analytic pseudo profiles can now be computed. For
the extreme ($D=10^{-7}$) Matos, Guzm\'{a}n and Nu\~{n}ez solution, these work
out to leading order in $r$ as
\begin{equation}
m_{\text{RC}}(r)=\frac{lr}{2}\approx10^{-6}r
\end{equation}%
\begin{equation}
m_{\text{lens}}(r)\approx\frac{l(l^{2}+l-4)r}{4(l^{2}-4)}\approx10^{-6}r
\end{equation}%
\begin{equation}
2(m_{\text{RC}}^{\prime}-m_{\text{lens}}^{\prime})\approx\frac{l(l^{2}%
-l-4)}{2(l^{2}-4)}\approx10^{-6}.
\end{equation}
The dimensionless parameter $\omega$ to all orders in $r$ with no restriction
on $D$ is
\begin{equation}
\omega(r)\approx\frac{2}{3}\frac{m_{\text{RC}}^{\prime}-m_{\text{lens}%
}^{\prime}}{2m_{\text{lens}}^{\prime}-m_{\text{RC}}^{\prime}}=\frac
{l(l^{2}-l-4)r^{2+l}-D(l^{3}+l^{2}-4l-4)}{3[D(l^{3}+l^{2}-4l-4)+l^{2}r^{2+l}%
]},
\end{equation}
which yields $\omega\approx3\times10^{5}$ for $D=10^{-7}$ within our chosen
range, $r\sim100-300$ $Kpc$. Note that if we straightaway put $D=0$ in
Eq.(21), we get $\omega(r)<-1$, conveying a completely wrong physical conclusion.

The pivotal result of the present article is that $D\geq10^{-7}$ and not
$D=0$, as discussed above. Of course, the lowest limit on $D$ is small and it
is quite tempting to set it exactly to zero. But the price for it is that one
gets a completely wrong picture of the halo. We have analyzed the model taking
into account only the lowest value of $D$. Similar analysis can be carried out
with other values of $D$ as well respecting the suggested lower limit. We can
say that, by and large, the conclusion of Matos, Guzm\'{a}n and Nu\~{n}ez
about non-Newtonian nature of the halo is right provided the restriction on
$D$ is maintained. With this restriction in place, their model can indeed be a
physically viable one. If combined measurements follow the pattern as
indicated in Eqs.(18)-(21), we would say that the model is observationally
supported. However, given the present uncertainties in observation, it is yet
too premature to say so.

We are deeply indebted to Guzel N. Kutdusova for her assistance at SSPA and
BSPU where the work was carried out.

\textbf{Figure captions:}

Fig.1. Plot of $\omega(r)$ vs $r$ in which $\omega$ is computed from either
Eqs.(6)-(8) or (21) with $l=10^{-6}$ and $D=10^{-7}$. The distance $r$ in
galactic halo region is taken in the range $100-300$ $Kpcs$. The non-Newtonian
values of $\omega$ are evident.

Fig.2. Plot of $\omega(r)$ vs $r$ in which $\omega$ is computed from either
Eqs.(6)-(8) or (21) with $l=10^{-6}$ and $D=10^{-8}$. The distance $r$ in
galactic halo region is taken in the range $100-300$ $Kpcs$. The figure
displays the transition behavior of $\omega$ as discussed in the text.

Fig.3. Plot of $\omega(r)$ vs $r$ in which $\omega$ is computed from either
Eqs.(6)-(8) or (21) with $l=10^{-6}$ and $D=10^{-9}$. The distance $r$ in
galactic halo region is taken in the range $100-300$ $Kpcs$. The values of
$\omega$ are negative indicating repulsion.

Fig.4. Plot of $\omega(r)$ vs $r$ in which $\omega$ is computed from either
Eqs.(6)-(8) or (21) with $l=10^{-6}$ and $D=0$. The distance $r$ in galactic
halo region is taken in the range $100-300$ $Kpcs$. The values of $\omega$ are
negative indicating repulsion, similar to that in Fig.3.

\textbf{References}

[1] S. Fay, Astron. Astrophys. \textbf{413}, 799 (2004)

[2] T. Matos, F.S. Guzm\'{a}n and D. Nu\~{n}ez, Phys. Rev. D \textbf{62},
061301 (2000)

[3] U. Nucamendi, M. Salgado and D. Sudarsky, Phys. Rev. D \textbf{63}, 125016 (2001)

[4] F. Rahaman, M. Kalam, A. DeBenedictis, A.A. Usmani and Saibal Ray, Mon.
Not. R. Astron. Soc. \textbf{389}, 27 (2008)

[5] T. Matos, F.S. Guzm\'{a}n, Class. Quantum Grav. \textbf{18}, 5055 (2001)

[6] M. Alcubierre, F.S. Guzm\'{a}n, T. Matos, D. Nu\~{n}ez, L.A.
Ure\~{n}a-L\'{o}pez and P. Wiederhold, Class. Quantum Grav. \textbf{19}, 5017 (2002)

[7] F.S. Guzm\'{a}n and L.A. Ure\~{n}a-L\'{o}pez, Phys. Rev. D \textbf{68},
024023 (2003)

[8] F.S. Guzm\'{a}n and L.A. Ure\~{n}a-L\'{o}pez, Astrophys. J. \textbf{645},
814 (2006)

[9] A. Bernal and F.S. Guzm\'{a}n, Phys. Rev. D \textbf{74}, 103002 (2006)

[10] S. Chandrasekhar, \textit{Mathematical Theory of Black Holes} (Oxford
University Press, 1983)

[11] A. Boriello and P. Salucci, Mon. Not. R. Astron. Soc. \textbf{323}, 285 (2001)

[12] S. Bharadwaj and S. Kar S, Phys. Rev. D \textbf{68}, 023516 (2003)

[13] D. Lynden-Bell, J. Katz and J. Bi\v{c}\'{a}k, Phys. Rev. D \textbf{75},
024040 (2007)

[14] K.K. Nandi, Y.Z. Zhang, R.G. Cai and A. Panchenko, Phys. Rev. D
\textbf{79}, 024011 (2009)

[15] K.K. Nandi, A. Islam and J. Evans, Phys. Rev. D \textbf{55}, 2497 (1997)

[16] K.K. Nandi, B. Bhattacharjee, S.M.K. Alam and J. Evans, Phys. Rev. D
\textbf{57}, 823 (1998)

[17] T. Faber and M. Visser, Mon. Not. R. Astron. Soc. \textbf{372}, 136 (2006)
\end{document}